%% file: main.tex
\documentclass[sigconf, balance=false]{acmart}

\usepackage{algorithm,algpseudocode}
\usepackage{booktabs} 
\usepackage{subcaption}
\usepackage{multirow}
\usepackage{bm}
\usepackage{amsthm}
\usepackage{amsmath}
\usepackage{threeparttable}
\usepackage{xspace}

\newcommand{\aw}{\emph{Agent Web}\xspace}
\newcommand{\method}{\textsc{AgentWebRec}\xspace}

\copyrightyear{2027}
\acmYear{2027}
\acmDOI{XXXXXXX.XXXXXXX}
\acmConference[KDD '27]{}{August 01--05, 2027}{San Jose, United States}
\acmBooktitle{KDD '27: The 33rd ACM SIGKDD International Conference on Knowledge Discovery \& Data Mining, August 01--05, 2027, San Jose, United States}
\acmPrice{15.00}
\acmISBN{978-1-4503-XXXX-X/18/06}

\begin{document}
\title{\method: Compact Evidence Fusion over the Agent Web for Personalized Recommendation}

\author{Haoran Qiang}
\affiliation{%
  \institution{MIIT Key Laboratory of Data and Decision Intelligence}
  \institution{Beihang University}
  \city{Beijing}
  \country{China}
}
\email{hr_qiang@buaa.edu.cn}

\author{Guannan Liu\textsuperscript{*}}
\affiliation{%
  \institution{MIIT Key Laboratory of Data and Decision Intelligence}
  \institution{Beihang University}
  \city{Beijing}
  \country{China}
}
\email{liugn@buaa.edu.cn}

\author{Liang Zhang}
\affiliation{%
  \institution{The Hong Kong University of Science and Technology (Guangzhou)}
  \city{Guangzhou}
  \country{China}
}
\email{liangzhang@hkust-gz.edu.cn}

\author{Junjie Wu}
\affiliation{%
  \institution{MIIT Key Laboratory of Data and Decision Intelligence}
  \institution{Beihang University}
  \city{Beijing}
  \country{China}
}
\email{wujj@buaa.edu.cn}

\input{abstract}




\maketitle

\input{introduction}

\input{preliminaries}

\input{method}

\input{experiment}

\input{relatedwork}

\input{conclusion}

\newpage
\bibliographystyle{ACM-Reference-Format}
\bibliography{references}
\newpage

\input{appendix}

\end{document}

%% file: abstract.tex
\begin{abstract}
LLM-based personal agents are emerging as persistent carriers of user semantics and intermediaries between users and recommendation platforms, maintaining richer user knowledge locally. As agents interact with one another, the conventional \textit{User--Platform} relation evolves into a \textit{User--Agent Web--Platform} information pathway, enabling distributed user-side information to complement item-side information. This new pathway, however, defies conventional recommendation: evidence is scattered across mutually opaque agents and reachable only through bounded queries, only a small portion of it is relevant to the current recommendation decision, and the responses returned by different agents are semantically heterogeneous. We therefore recast recommendation over the \aw{} as a \emph{task-time evidence acquisition and fusion} problem under a finite evidence budget by deciding what to ask and what to keep, rather than learning from aggregated data. We propose \method, a user-agent-oriented framework that progressively acquires and fuses distributed evidence for each user-item decision while keeping underlying agent memories local. It grounds each decision in platform-provided item semantics and task-relevant evidence from the target user agent's private memory, and conditionally queries neighboring user agents for complementary preference patterns when local evidence is insufficient. Experiments on four InstructRec datasets show that \method consistently outperforms baseline recommenders, and ablations verify that the evidence layers contribute complementary gains. 
\end{abstract}

%% file: introduction.tex
\section{Introduction}

\begin{figure}
    \centering
    \resizebox{0.98\linewidth}{!}{
        \includegraphics{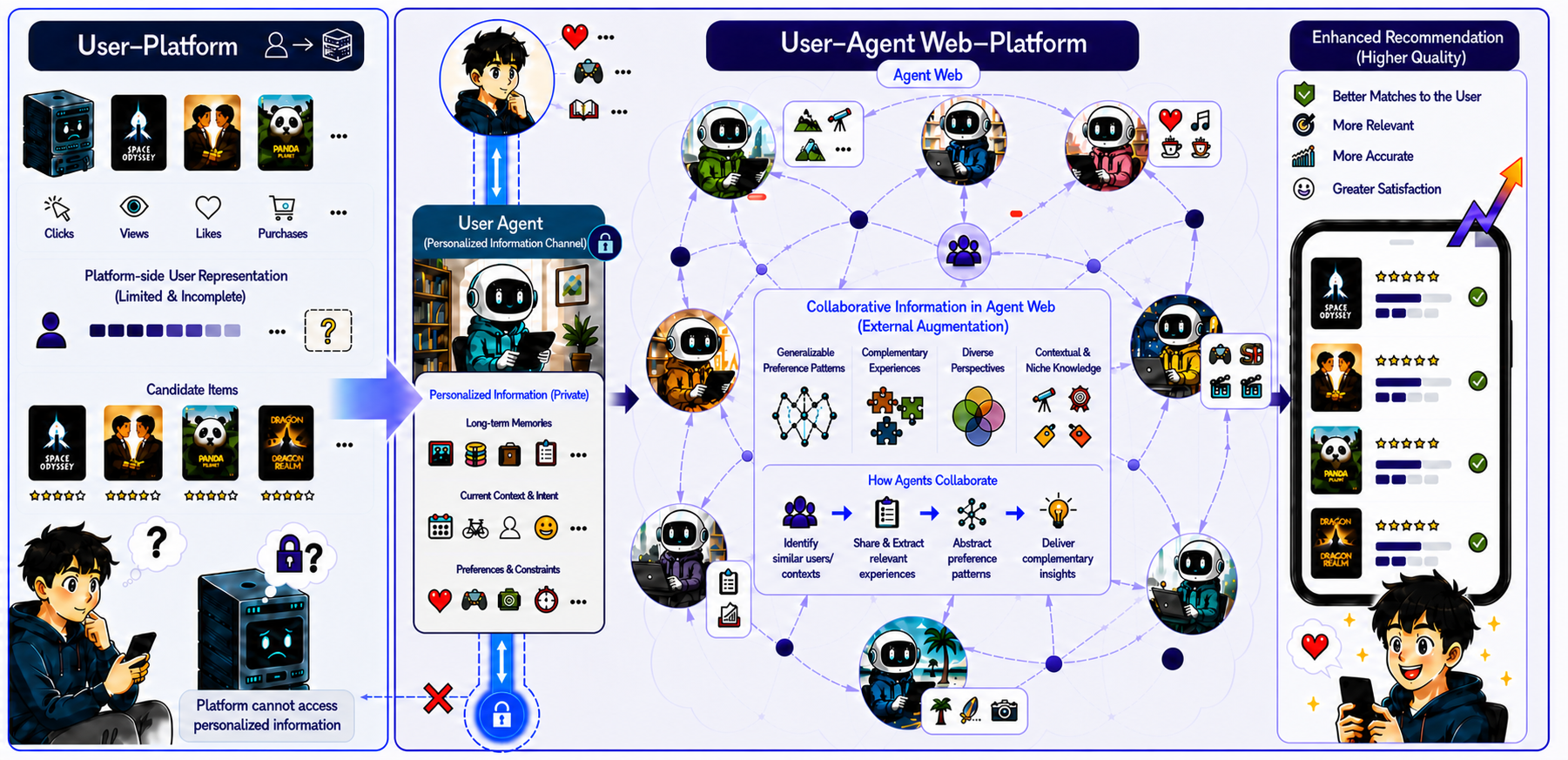}
    }
    \caption{LLM-based user agents reshape personalized recommendation: the \emph{User--Platform} relation evolves into a \emph{User--Agent Web--Platform} paradigm.}
    \label{fig:Intro_Fig}
    \vspace{-6pt}
\end{figure}

Recommendation systems have long served as a core infrastructure by which individual users discover relevant information on the Web, proactively surfacing items aligned with personal preferences instead of requiring users to search an ever-growing space themselves. Classical recommendation methods predominantly operate on direct user–platform interactions, from which they learn user and item representations to model user preferences~\cite{Koren2009Matrix,He2017NCF,Kang2018SASRec,He2020Light}. 
Although this platform-centric setting has enabled substantial progress in recommendation, 
interaction signals remain ambiguous. Because observed behaviors reveal only the users' actions but not the underlying intentions, aggregating such heterogeneous signals into user representations could emphasize recurring patterns while obscuring transient or evolving preferences.
Consequently, similar behavioral histories would correspond to different individual preferences, while the same user exhibits distinct needs under different contexts~\cite{Zhang2023InstructRec,Lin2024ReLLa}.

With the development of LLMs in recent years, LLM-based personal agents are beginning to change this interaction model. 
Consumer-oriented agent products, such as Alibaba's Qwen and Amazon's Alexa+, combine historical user information and current context to infer intentions and preferences~\cite{alibaba2026qwenapp, amazon2025alexa}. On this basis, an agent act on the user's behalf to interact with the platform by formulating searching for specific items, examining ranked results and comparing candidates. 
In contrast to conventional user-platform interactions, where user actions are observed as isolated stateless events, an agent maintains the evolving states across requests.
Meanwhile, memory-oriented agent systems show that an LLM agent can organize and retrieve long-term user-specific information \cite{Packer2023MemGPT,Park2023GenerativeAgents,wei2025secondme}. 
A user agent plays two roles in this new interaction model. As a persistent carrier of user semantics, it accumulates the user's preferences, feedback, and context across sessions and services, 
maintaining a local user profile richer than what any single platform observes. As an intermediary, it stands between the user and platforms, interpreting the user's intentions and acquiring information on the user's behalf. However, richer local knowledge does not guarantee sufficient task-time evidence: when prior experience offers little analogous signal for the candidate item, an isolated agent lacks the reusable preference patterns needed to ground its decision, and must look beyond its own memory.


Emerging agent-native communities provide a basis for connecting distributed information. Within this setting, when private memory lacks available information for a recommendation task, a user agent can discover and collaborate with agents that provide relevant information \cite{secondmeNetwork2026,mochat2026}.
In a macro view, task-oriented collaborations between user agents constitute the edges of the Agent Web \cite{chen2025internet,yang2025agenticweb}. Crucially, such collaboration only exposes task-level capabilities and responses with the internal states unseen: protocols such as A2A make agents discoverable while leaving their internal memories and reasoning states opaque \cite{a2a2026spec}.
Agent-native communities are not limited to interactions among user agents: recommendation platforms also participate in the \aw{} through functional agents that expose item-side information~\cite{chen2025internet,a2a2026spec}. As shown in Figure~\ref{fig:Intro_Fig}, the conventional \emph{User--Platform} relation is thereby transformed into a \emph{User--Agent Web--Platform} paradigm, where the user agent interprets preferences from local semantic memory and integrates task-relevant evidence acquired across the \aw.

Existing recommendation research does not directly address this configuration. LLM-based recommendation methods use LLMs to improve the representation of user intentions, item semantics, interaction sequences, and recommendation instructions \cite{Geng2022P5,hou2024llmrank,Li2023Recformer,Wu2023LLMRecSurvey,Zhang2023InstructRec}. Agent-based approaches further introduce planning, tool use, user simulation, reflection, feedback loops, and multi-agent coordination into recommendation \cite{Cai2024AFL,Huang2023InteRecAgent,Wang2023RecMind,Wang2024MACRec,Zhang2023Agent4Rec,Zhang2023AgentCF}. These studies demonstrate the potential of agent reasoning and memory, but generally organize agents within a platform-controlled workflow or allow the platform to construct and update reusable agent states. Meanwhile, research on the Internet of Agents and the Agentic Web studies general-purpose agent discovery, communication, and collaboration \cite{chen2025internet,yang2025agenticweb}, but does not specify how independently maintained user-side information should be combined with platform-side information for a specific recommendation decision.  

This gap changes the central problem from aggregating more user information to eliciting the information required by the current decision. We therefore ask: \textit{How can a user agent construct task-specific queries that acquire and fuse useful information from distributed and opaque sources for a current recommendation task?} This question presents three challenges. 
First, the relevant information is distributed across participants whose internal states are mutually inaccessible. A requester must express what information is needed without inspecting the underlying evidence in advance. 
Second, richer information is not necessarily more useful. A user agent maintains extensive long-term information, but only a small fraction may be relevant to the current decision context. Unconstrained acquisition can introduce irrelevant information and dilute the target user's actual preference. 
Third, the returned responses are heterogeneous. Platform knowledge, agent local information, and external agent feedback describe different aspects of the task and may differ substantially in semantic perspective. Directly concatenating them does not guarantee that they support the decision.

To investigate this problem, we propose \textsc{\method}, a user agent oriented recommendation framework that formulates recommendation as task-time information acquisition and fusion over the Agent Web. \method treats task-specific queries as the interface between distributed information sources and the current decision. Instead of transferring complete memories or reconstructing a global user profile, each participant interprets the information query locally and returns a compact response. The target user agent then organizes the acquired information around the current user-item context, allowing user-side information to complement the platform representation and external experience to provide additional support when appropriate. 

We evaluate \method on four domains from the InstructRec benchmark and compare it with conventional recommendation methods. \method achieves consistently strong performance across the evaluated datasets. More analyses further demonstrate that target-agent information and external-agent experience provide complementary benefits, while task-oriented information acquisition avoids indiscriminately introducing all available agent information.
Together, the contributions are summarized as follows:

\begin{itemize}
    \item We introduce recommendation over an \aw composed of user agents, which mediates interactions between users and the platform while agent memories remain distributed and local.
    \item We propose \method, which selectively acquires distributed information and progressively fuses it into task-specific evidence without centralizing agent memories.
    \item Experiments on four InstructRec datasets demonstrate the effectiveness of \method against baselines for personalized recommendation.
\end{itemize}

%% file: preliminaries.tex
\section{Problem Definition}
\label{sec:pre}

We consider the problem as recommendation over the Agent Web. 
Specifically, an \aw is represented as $\mathcal G=(\mathcal A,\mathcal E)$, where $\mathcal A$ is a set of LLM-based agents and $\mathcal E\subseteq\mathcal A\times\mathcal A$ indicates the established task-conditioned interactions among agents. 

Each agent $A_a\in\mathcal A$ interacts directly with its neighbors $\mathcal N_a,$ which share similar features \cite{mehdizadeh2025homophily,hashemi2026empirical}.
The Agent Web therefore provides a distributed information environment in which agents collaborate through the query-response interactions to obtain complementary information from one another.
Given a query $q$, the agent $A_a$ processes the query using its memory $\mathcal M_a$ and returns a response: $r=A_a(q;\mathcal M_a).$

For recommendation scenario, we instantiate the \aw  with a platform agent $A_P$ and a collection of user agents $\mathcal A_U=\{A_u:u\in\mathcal U\}$, yielding the complete agent set $\mathcal A=\{A_P\}\cup\mathcal A_U$.
The platform agent maintains an item memory $\mathcal M_P=\{m_j^P:j\in\mathcal I\}$, where each record 
$m_j^P=(j,d_j)$ contains the semantic description $d_j$ of item $j\in\mathcal I$. 
Each user agent $A_u$ maintains a personalized memory $\mathcal M_u=\left(\phi_u,\{m_{u,k}\}_{k=1}^{n_u}\right),$ where $\phi_u$ denotes the user persona, $m_{u,k}$ denotes the interaction record, and $n_u$ is the number of records. Each interaction record 
$m_{u,k}=(i_{u,k},d_{u,k},\ell_{u,k},\tau_{u,k})$ consists of a historically interacted item $i_{u,k}$, its description $d_{u,k}$, the user's feedback or review $\ell_{u,k}$, and the interaction time $\tau_{u,k}$.

A recommendation task, specified by a triple $t=(u,i,\tau)$ with target user $u\in\mathcal U$, candidate item $i\in\mathcal I$, and decision time $\tau$, is to predict the preference $Y_t$ of user $u$ toward item $i$ at time $\tau$.
Solving task $t$ in the \aw faces two constraints. First, task-relevant information is distributed across agents whose memories remain mutually isolated, so centralized training over 
aggregated data is infeasible. Second, agent memories accumulate interactions from diverse contexts; indiscriminately incorporating unrelated records or weakly aligned agent responses dilutes the preference signals relevant to the current task~\cite{bai2024longbench,liu2024lost,yoran2024robust}. 
Effective recommendation therefore requires concentrating dispersed information into compact evidence that preserves task-relevant semantics while excluding distracting content.

We accordingly formulate recommendation over the \aw as a task-time information acquisition and evidence fusion problem. For task $t$, let $\mathcal R_t$ denote the information acquired from 
distributed agent memories through query--response interactions, and let $Fuse(\cdot)$ denote the function that identifies task-relevant content within $\mathcal R_t$ and aligns heterogeneous sources into the evidence for prediction. The problem is to determine which information to acquire as $\mathcal R_t$ and how to fuse it, such that the resulting prediction $\hat Y_t=Fuse(\mathcal R_t)$ approximates 
the true preference $Y_t$.


%% file: method.tex
\section{Methodology}
\label{sec:method}

\begin{figure*}
    \centering
    \resizebox{0.85\linewidth}{!}{
        \includegraphics{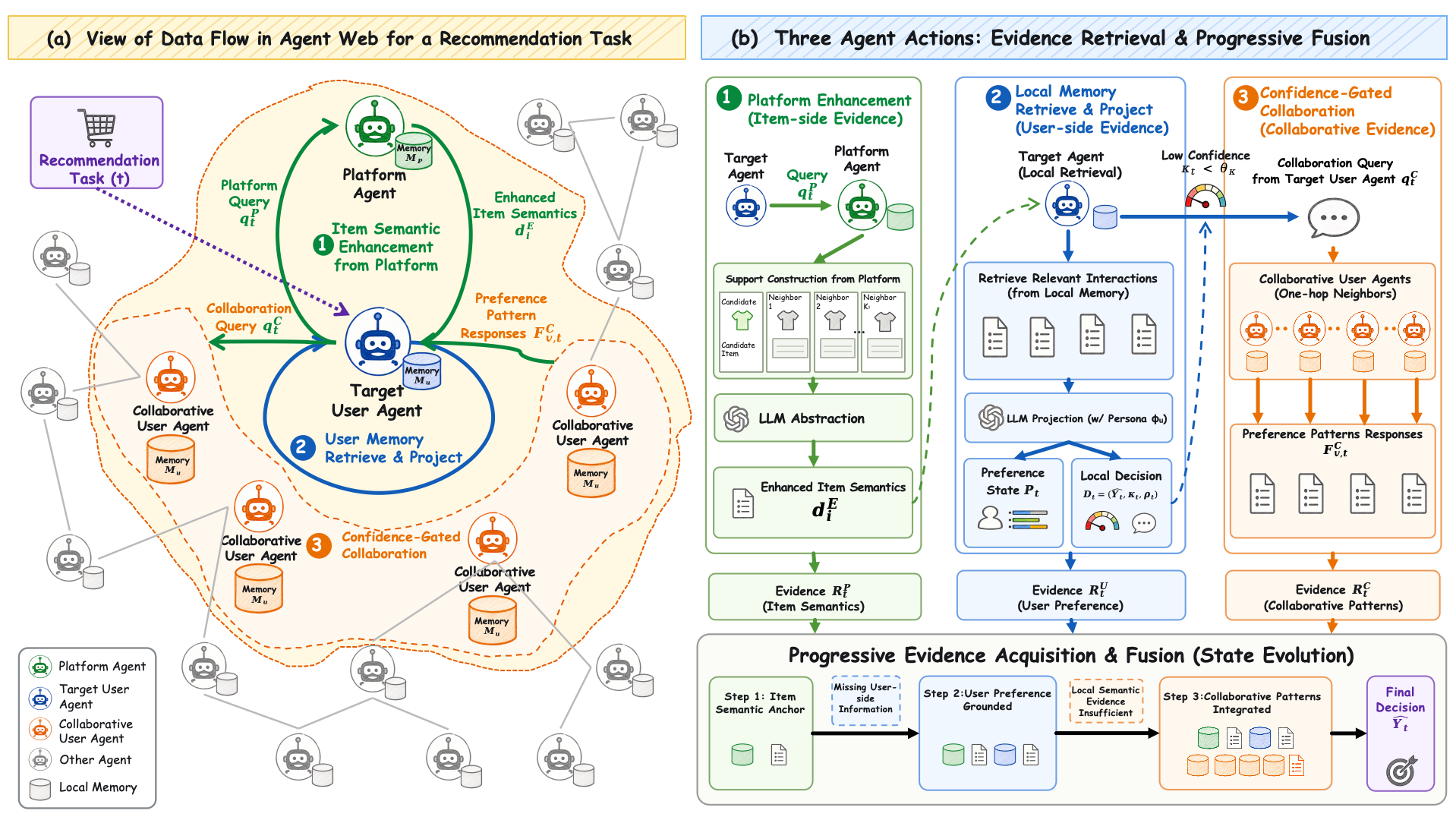}
    }
    \caption{Overview of \method for progressive information acquisition and evidence fusion over the \aw. (a) For a recommendation task $t$, the target user agent acquires item semantics from the platform, retrieves user preferences from local memory, and conditionally queries neighboring user agents for generalizable patterns. (b) The resulting platform, user, and collaborative information is progressively fused as evidence to produce the final recommendation $\hat{Y}_t$.}
    \label{fig:ARec_Framework}
\end{figure*}

\subsection{Overview of \method}
\label{subsec:method_overview}

To address the two coupled problems formulated above, we propose \method to selectively acquire distributed information and progressively fuse it into compact evidence for each recommendation task.
\method instantiates \(Fuse(\cdot)\) as a progressive process that interleaves task-time information acquisition with evidence fusion over the Agent Web. 

As shown in the Figure~\ref{fig:ARec_Framework}(a), the target user agent performs the recommendation task by interacting with the platform agent, its own local memory, and its neighboring user agents.
The query-response data flow enables interaction among agents whose internal states are mutually inaccessible. A query and a response are generated autonomously by an agent according to its current internal state without knowing the external states from other collaborators. As a result, no centralized orchestrator is required to maintain a global view of the participating agents in the \aw.

Within this decentralized interaction interface, the three agent actions in Figure~\ref{fig:ARec_Framework}(b) progressively identify information relevant to the current decision. The platform query \(q_t^P\) elicits enhanced item semantics \(d_i^E\) as \(\mathcal R_t^P\), which establishes a preference-relevant semantic anchor for the candidate item. This anchor determines the scope of local retrieval, allowing the target user to select relevant interactions from its own memory as \(\mathcal R_t^U\). The retrieved information is projected into a task-specific preference state \(P_t\) and an initial decision \(D_t\). 
When the semantic evidence from target user agent is insufficient, the unresolved decision is transformed into a collaboration query \(q_t^C\), which elicits generalizable preference patterns evidence from neighboring agents as \(\mathcal R_t^C\). These patterns complement the limited coverage of the target user's private experience and strengthen the reliability of the resulting decision.
Information acquisition is therefore controlled by the evolving task state, which respects to the relation $\mathcal R_t = \mathcal R_t^P \cup \mathcal R_t^U \cup \mathcal R_t^C,$  where \(\mathcal R_t^C=\varnothing\) when collaboration is not activated.  
Together, \method constructs compact, task-oriented evidence without exposing agent-internal states, using query-response interactions to elicit decision-relevant information and progressive state updates to reconcile the heterogeneous semantics of different sources.

\subsection{Querying the Platform Agent for Item Semantic Enhancement}
\label{subsec:platform_enhancement}

In recommendation over the \aw, a candidate item initially carries only a limited semantic profile. Thus, anchoring the task on such a sparse representation introduces ambiguity into subsequent memory retrieval 
and preference reasoning. \method therefore first queries the platform agent for a semantic enhancement before $A_u$ performs personalized evidence acquisition.

Specifically, after receiving task \(t\), the target user agent sends a query \(q_t^P = t(i)\)   to \(A_P\), specifying the candidate item \(i\) whose semantics need to be enhanced.
Based on the item descriptions, \(A_P\) retrieves the semantic neighbors of candidate item \(i\):
\begin{equation}
    \mathcal J_t(i)
    =
    \operatorname{TopK}_{j\in\mathcal I\setminus\{i\}}^{K_I-1}
    \operatorname{sim}\!\left(
        g_{\mathrm{emb}}(d_i),
        g_{\mathrm{emb}}(d_j)
    \right),
\end{equation}
where \(g_{\mathrm{emb}}(\cdot)\) denotes a semantic encoder that maps an item description to its profile embedding, and \(\operatorname{sim}(\cdot,\cdot)\) denotes cosine similarity between normalized embeddings.
The set \(\mathcal J_t(i)\) contains the \(K_I-1\) items most semantically similar to \(i\). Together with the candidate item, these neighbors form a platform-side support set of size \(K_I\) for subsequent semantic abstraction.


The platform agent abstracts the relevant item information into enhanced item semantics with LLM:
\begin{equation}
\begin{aligned}
    d_i^E
    =
    A_P(q_t^P;\mathcal M_P)
    =
    \operatorname{AbstractLLM}\!\left(
    m_{i}^P,
    \{m_{j}^P:j\in\mathcal J_t(i)\}
    \right).
\end{aligned}
\end{equation}
The response \(d_i^E\) preserves the original description \(d_i\), while enriching them with similar semantics distilled from the retrieved semantic neighbors. 
The support construction follows a raw-preserving enhancement principle with the prompt design: semantic neighbors supplement incomplete item semantics, while the candidate record prevents the enhancement from drifting away from the item. 
The concrete prompt used to instantiate \(\operatorname{AbstractLLM}\) is provided in \textbf{Prompt~\ref{fig:AbstractLLM}} in Appendix~\ref{sec:llm_prompts}. 
From the perspective of $A_u$, only this compact query-response interaction is acquired as platform-side information: $\mathcal R_t^P = \{(q_t^P,d_i^E)\}$ without the original information of the related items.

\subsection{Retrieving and Projecting Private Memory}
\label{subsec:user_memory_retrieval}

The enhanced candidate semantics must be grounded in the target user's private experience before a personalized preference can be inferred. However, $\mathcal M_u$ contains heterogeneous interactions whose relevance varies across candidate items and decision times. Reasoning over the complete memory would therefore introduce unrelated or outdated information into the task-time evidence. 

After receiving the enhanced item semantics $d_i^E$ responded  from the platform agent , \(A_u\)  uses the resulting embedding \(g_{\mathrm{emb}}(d_i^E)\) as a semantic anchor for candidate-conditioned retrieval from its private memory \(\mathcal M_u\). This local matching process identifies historical interactions relevant to the candidate item.
For each historical interaction $m_{u,k}$, its semantic relevance to the current recommendation task is measured by
\[
    \omega_{t,k}^{\mathrm{sem}}
    =
    \operatorname{sim}\!\left(
    g_{\mathrm{emb}}(d_i^E),
    g_{\mathrm{emb}}(d_{u,k})
    \right).
\]
The temporal relevance is measured by 
 \[\omega_{t,k}^{\mathrm{temp}}
    =
    \exp\!\left[
    -\gamma(\tau-\tau_{u,k})
    \right], 
 \] where $\gamma\geq 0$ controls the temporal decay rate. 
 The two relevance signals are combined as
\begin{equation}
    \omega_{t,k}
    =
    \beta\omega_{t,k}^{\mathrm{sem}}
    +
    (1-\beta)\omega_{t,k}^{\mathrm{temp}},
\end{equation}
where $\beta\in[0,1]$ balances candidate-specific semantic relevance and temporal recency. 
The semantic component suppresses interactions unrelated to candidate item $i$, whereas the temporal component accounts for changes in the user's preferences over time.


$A_u$ selects the memory subset
\begin{equation}
    \widehat{\mathcal M}_{u,t}
    \subseteq \{m_{u,k}\}_{k=1}^{n_u},
    \qquad
    |\widehat{\mathcal M}_{u,t}| = \min(K_M, n_u),
\end{equation}
containing the records with the $K_M$ largest relevance scores
$\omega_{t,k}$.

The selected records induce the target-user information
\begin{equation}
    \mathcal R_t^U
    =
    \{
    h_{t,k}^U:
    m_{u,k}\in\widehat{\mathcal M}_{u,t}
    \},
\end{equation}
where $h_{t,k}^U$ denotes the task-specific evidence induced by interaction record $m_{u,k}$. Thus, only a bounded subset of the private memory enters the evidence acquired for the current recommendation task.

To align the filtered evidence retrieved from $\mathcal M_u,$ $A_u$ constructs a candidate-conditioned temporal preference state, with LLM projecting the user persona and retrieved evidence onto the enhanced candidate semantics:
\begin{equation}
    P_t
    =
    \operatorname{ProjectLLM}\!\left(
    \phi_u,
    \mathcal R_t^U
    \right).
\end{equation}
\(\operatorname{ProjectLLM}\) identifies preference patterns supported by the retrieved interactions and relevant to the enhanced candidate semantics, and organizes them into the compact preference state \(P_t\), which serves for subsequent actions. The concrete prompt used to instantiate \(\operatorname{ProjectLLM}\) is provided in \textbf{Prompt~\ref{fig:ProjectLLM}} in Appendix~\ref{sec:llm_prompts}.

With this task-aligned projection, $A_u$ matches the projected preference with the enhanced candidate semantics and generates a decision with LLM:
\begin{equation}
    D_t
    =
    (\hat{Y_t},\kappa_t,\rho_t)
    =
    \operatorname{MatchLLM}\!\left(
    P_t,d_i^E
    \right).
\end{equation}
\(\operatorname{MatchLLM}\) evaluates the semantic alignment between the projected preference and the enhanced candidate, producing the local decision \(D_t\). Here, \(\hat Y_t\) is the preference prediction, \(\kappa_t\) measures whether the available personalized evidence is sufficient to support a reliable decision, and \(\rho_t\) is the task-specific rationale. The concrete prompt used to instantiate \(\operatorname{MatchLLM}\) is provided in \textbf{Prompt~\ref{fig:MatchLLM}} in Appendix~\ref{sec:llm_prompts}.

\subsection{Confidence-Gated Collaboration in the Agent Web}
\label{subsec:confidence_collaboration}

Users with similar historical behaviors are more likely to exhibit shared recommendation-relevant patterns. The \aw follows this preference homophily principle through neighboring user agents, each maintaining private experience and returning query-conditioned pattern summaries. Low confidence in the decision indicates that the semantic evidence available in the target user's memory is insufficient to characterize the current recommendation context. \method therefore activates collaboration utilizing their generalizable preference patterns to provide complementary support for the decision.

Specifically, collaboration is activated when $\kappa_t<\theta_\kappa$, where $\theta_\kappa$ is the confidence threshold. The target user agent abstracts its unresolved decision context into a collaboration query with LLM:
\begin{equation}
    q_t^C
    =
    \operatorname{QueryLLM}\!\left(
    d_i^E,P_t,\rho_t
    \right).
\end{equation}
The query $q_t^C$ specifies the complementary information required from collaborative agents establishing a shared semantic plane onto which their interaction histories are projected.
The concrete prompt used to instantiate \(\operatorname{QueryLLM}\) is provided in \textbf{Prompt~\ref{fig:QueryLLM}} in Appendix~\ref{sec:llm_prompts}.

The collaborative agents are the one-hop user-agent neighbors of $A_u$ in the \aw.
Each collaborator $A_v\in\mathcal N_u$ receives the common query 
$q_t^C$ and retrieves its own task-relevant interaction memory
\begin{equation}
    \widehat{\mathcal M}_{v,t}
    \subseteq \{m_{v,k}\}_{k=1}^{n_v},
    \qquad
    |\widehat{\mathcal M}_{v,t}| = \min(K_N, n_v),
\end{equation}
containing the records with the $K_N$ largest query--memory 
relevance scores 
$\operatorname{sim}\!\bigl(
g_{\mathrm{emb}}(q_t^C),\, g_{\mathrm{emb}}(d_{v,k})\bigr)$, 
where $K_N$ upper-bounds the number of records each collaborator 
contributes to task $t$.

Conditioned on $q_t^C$, $A_v$ abstract the retrieved experience into a compact preference-pattern response with LLM:
\begin{equation}
\begin{aligned}
    F_{v,t}^C
    =
    A_v(q_t^C;\mathcal M_v) 
    =
    \operatorname{PatternLLM}\!\left(
    q_t^C,
    \phi_v,
    \widehat{\mathcal M}_{v,t}
    \right).
\end{aligned}
\end{equation}
\(\operatorname{PatternLLM}\) identifies preference regularities that are grounded in the collaborator's relevant experiences and responsive to the information need expressed by \(q_t^C\). The resulting \(F_{v,t}^C\) provides a generalizable preference pattern, while the supporting original interaction records remain local. The concrete prompt used to instantiate \(\operatorname{PatternLLM}\) is provided in \textbf{Prompt~\ref{fig:PatternLLM}} in Appendix~\ref{sec:llm_prompts}.

These responses induce the collaborative evidence
\begin{equation}
  \mathcal R_t^C
    =
    \{
    (A_v,q_t^C,F_{v,t}^C):
    A_v\in\mathcal N_u
    \}  
\end{equation}
When collaboration is activated, $A_u$ incorporates the returned preference patterns into the origin decision with LLM:
\begin{equation}
    (\hat{Y_t},\kappa_t',\rho_t')
    =
    \operatorname{RefineLLM}\!\left(
    D_t,
    \mathcal R_t^C
    \right).
\end{equation}
\(\operatorname{RefineLLM}\) evaluates the returned preference patterns together with the original local decision, using signals that recur across collaborators as generalizable pattern information and incorporating only those that provide relevant complementary support. It preserves the local judgment when the collaborative information is irrelevant or conflicts with the target user's personalized evidence, and otherwise updates the preference prediction, confidence, and rationale to reflect the additional collaborative evidence.
The concrete prompt used to instantiate \(\operatorname{RefineLLM}\) is provided in \textbf{Prompt~\ref{fig:RefineLLM}} in Appendix~\ref{sec:llm_prompts}.


%% file: experiment.tex
\section{Experiments}

We conduct experiments to evaluate whether distributed user agents can autonomously infer task-specific preferences from decentralized user evidence and construct effective recommendation signals. The experiments are organized to answer the following four research questions:

\textbf{RQ1 (Overall Performance):} How does \method perform compared with conventional centralized recommendations and existing LLM/agent-based recommendation methods?

\textbf{RQ2 (Agent Actions Contribution):} How do the key agent actions in \method contribute to recommendation effectiveness?

\textbf{RQ3 (Parameter Sensitivity):} How do key task-time parameters affect the recommendation performance of \method?

\textbf{RQ4 (Collaboration Structure):} How does the construction of user agent edges in the Agent Web affect recommendation performance?





\subsection{Experimental Setup}

\subsubsection{Datasets}

We evaluate \method on four domains from the InstructRec benchmark~\cite{Zhang2023InstructRec}: Books, Goodreads, MovieTV, and Yelp. Each dataset provides timestamped user-item interactions, user persona descriptions, item profiles, and review texts. Table~\ref{tab:dataset_stats} summarizes the statistics of the four datasets.

Following common experimental settings in agent-based recommendation~\cite{Zhang2023AgentCF, liu2025agentcfpp,Chen2026MemRec}, we order each user's interactions temporally, use the earlier records as warm-up history and the last interaction as the test target. For evaluation, we randomly sample 1,000 users and rank each target item against nine sampled negatives.
User agents operate continuously as autonomous participants in the Agent Web without activation by the users' instruction. We therefore provide no instruction and require the target user agent to infer the user's relevant preference.
We report Hit Rate (\textbf{H@K}) and NDCG (\textbf{N@K}) for \(K\in\{1,3,5\}\).
To reduce randomness, we report the average results of three repeating tests.
By default, \method enhances each candidate with five semantic neighbors, retrieves up to five private-memory interactions for the target and each collaborator, and activates at most three collaborators over item-co-occurrence edges under a confidence threshold of 0.7.

\begin{table}[t]
\centering
\caption{Statistics of the InstructRec datasets used in our experiments.}
\label{tab:dataset_stats}
\small
\setlength{\tabcolsep}{2.2pt}
\resizebox{0.38\textwidth}{!}{
\begin{tabular}{@{}lrrrrr@{}}
\toprule
\textbf{Dataset} & \textbf{Users} & \textbf{Items} & \textbf{Inter.} & \textbf{Dens. (\%)} & \textbf{Inter./U} \\
\midrule
Books     & 7,377  & 120,925 & 207,759 & 0.0233 & 28.16 \\
Goodreads & 11,734 & 57,364  & 618,330 & 0.0919 & 52.70 \\
MovieTV   & 5,649  & 28,987  & 79,737  & 0.0487 & 14.12 \\
Yelp      & 2,950  & 31,636  & 63,142 & 0.0677 & 21.40 \\
\bottomrule
\end{tabular}
}
\vspace{-6pt}
\end{table}

\vspace{-6pt}
\subsubsection{Baselines}
We compare \method with two groups of baselines. The first group contains conventional centralized recommendation models: \textbf{SASRec}~\cite{Kang2018SASRec} and \textbf{LightGCN}~\cite{He2020Light}. SASRec represents sequential recommendation based on self-attention over user behavior sequences, while LightGCN represents graph-based collaborative filtering over centralized user-item interactions.
The second group contains LLM-based or agent-based recommendation methods adapted to the same InstructRec evaluation protocol~\cite{Zhang2023InstructRec}: \textbf{LLMRank}~\cite{hou2024llmrank}, \textbf{AFL}~\cite{Cai2024AFL}, \textbf{AgentCF}~\cite{Zhang2023AgentCF}, \textbf{AgentCF++}~\cite{liu2025agentcfpp}, and \textbf{MemRec}~\cite{Chen2026MemRec}. LLMRank directly uses an LLM as a zero-shot listwise ranker over user histories and candidate items. AFL combines a sequential recommendation prior with LLM-based feedback-loop reranking. AgentCF learns user and item agents through feedback-driven language updates. AgentCF++ extends AgentCF with shared group memory for collaborative reasoning. MemRec retrieves and prunes user/item memories before LLM reranking.
For controlled comparison, all applicable methods use Qwen2.5-7B-Instruct as the LLM backend and all-MiniLM-L6-v2 as the semantic encoder.

\subsection{Main Results (RQ1)}
\label{subsec:main_results}

We compare \method with centralized representation-based recommendation methods to assess whether it can exceed platform-centric recommendation under the \textit{User--Agent Web--Platform} setting.
Table~\ref{tab:main_results} reports the main comparison results. 
As seen, \method achieves the best results across all reported metrics on the four datasets. Its advantages are particularly clear on H@1 and NDCG for Books, Goodreads, and MovieTV, indicating that the acquired information helps the target agent distinguish the ground-truth item at the highest ranking positions. The consistent gains in H@3, H@5, and NDCG further show that the improvement is not limited to a single top-ranked prediction.

LightGCN and SASRec primarily learn from structural regularities in user-item interactions, so their support becomes limited in the sparse datasets whose preference inference is mainly from the semantic textual content. 
LLMRank and AFL rely on LLMs to reason over the semantic content of user-item histories. However, irrelevant information can obscure useful evidence without task-specific filtering while the absence of interactions among user agents prevent them from capturing generalized collaborative patterns.
AgentCF and AgentCF++ user collaborative memory propagation to capture generalizable patterns across user agents. However, they encode user preference and collaborative information into persist profiles, mixing information from different sources before decision. \method instead tightly couples its acquisition and fusion actions, with each action guided by the state produced by the preceding one. 
This progressive design integrates task-relevant information from heterogeneous sources into reliable evidence, introducing collaborative patterns only when needed without diluting the target user's personalized information.
MemRec is generally the strongest baseline, as it curates and synthesizes target-user information with collaborative signals into a compact reasoning context. However, its reliance on an explicit user instruction weakens its ability. \method instead uses the enhanced candidate semantics to retrieve relevant interactions from the target user's private memory and projects them with the user persona into a task-specific preference state, thereby identifying the user's current preference without a predefined instruction.

The performance can be understood from two dimensions: interaction availability and semantic understanding. LightGCN benefits from dense interactions on Goodreads but weakens when sparse overlap limits structural support. AgentCF and AgentCF++ do not consistently benefit from richer histories. Goodreads contains longer and more diverse reading histories, where additional interactions span multiple preference facets and therefore provide less concentrated support for a particular recommendation task.
\method remains robust by acquiring a well-calibrated information set: sufficient semantic evidence under sparse local support and collaborative patterns only when necessary.

\begin{table*}[t]
\centering
\caption{Main recommendation performance on the four InstructRec datasets. The best and second-best results are highlighted in \textbf{bold} and \underline{underlined}, respectively.}
\label{tab:main_results}
\scriptsize
\resizebox{0.88\textwidth}{!}{
\begin{tabular}{lccccc@{\hspace{0.5cm}}ccccc}
\toprule
& \multicolumn{5}{c}{\textbf{Books}}
& \multicolumn{5}{c}{\textbf{Goodreads}} \\
\cmidrule(lr){2-6}\cmidrule(lr){7-11}
\textbf{Model}
& \textbf{H@1} & \textbf{H@3} & \textbf{N@3} & \textbf{H@5} & \textbf{N@5}
& \textbf{H@1} & \textbf{H@3} & \textbf{N@3} & \textbf{H@5} & \textbf{N@5} \\
\midrule
LightGCN    & 0.0490  & 0.1490  & 0.1037 & 0.5410  & 0.2625 & 0.2820  & 0.4970  & 0.4060 & \underline{0.6790}  & \underline{0.4811} \\
SASRec      & 0.1370  & 0.3820  & 0.2765 & 0.5660  & 0.3513 & 0.1890  & 0.3730  & 0.2961 & 0.5060  & 0.3506 \\
LLMRank     & 0.2453  & 0.4243  & 0.3465 & 0.5853  & 0.4126 & 0.1827  & 0.4363  & 0.3276 & 0.6250  & 0.4055 \\
AFL         & 0.2080 & 0.4280 & 0.3346 & 0.5720  & 0.3934 & \underline{0.2893}  & \underline{0.4973}  & \underline{0.4088} & 0.6433  & 0.4691 \\
AgentCF     & \underline{0.4850} & \underline{0.7227} & \underline{0.6230} & \underline{0.8403} & \underline{0.6718} & 0.1110 & 0.2720 & 0.2037 & 0.3710 & 0.2442 \\
AgentCF++   & 0.2227  & 0.4360  & 0.3433 & 0.6040  & 0.4121 & 0.1730  & 0.4273  & 0.3177 & 0.5967  & 0.3870 \\
MemRec      & 0.3020 & 0.5690 & 0.4571 & 0.7290 & 0.5225 & 0.1870 & 0.4660 & 0.3471 & 0.6380 & 0.4173 \\
\method
            & \textbf{0.5250} & \textbf{0.8173} & \textbf{0.6996} & \textbf{0.9327} & \textbf{0.7483}
            & \textbf{0.6860} & \textbf{0.9460} & \textbf{0.8480} & \textbf{0.9910} & \textbf{0.8668} \\
\midrule
& \multicolumn{5}{c}{\textbf{MovieTV}}
& \multicolumn{5}{c}{\textbf{Yelp}} \\
\cmidrule(lr){2-6}\cmidrule(lr){7-11}
\textbf{Model}
& \textbf{H@1} & \textbf{H@3} & \textbf{N@3} & \textbf{H@5} & \textbf{N@5}
& \textbf{H@1} & \textbf{H@3} & \textbf{N@3} & \textbf{H@5} & \textbf{N@5} \\
\midrule
LightGCN    & 0.0730  & 0.2590  & 0.1767 & 0.5060  & 0.2772 & 0.0860  & 0.2870  & 0.1979 & 0.5120  & 0.2894 \\
SASRec      & 0.1010  & 0.3100  & 0.2204 & 0.5240  & 0.3079 & 0.0380  & 0.2370  & 0.1481 & 0.5020  & 0.2568 \\
LLMRank     & 0.2683  & 0.4890  & 0.3933 & 0.6440  & 0.4568 & 0.1337  & 0.3483  & 0.2559 & 0.5380  & 0.3339 \\
AFL         & 0.2563  & 0.5067  & 0.3982 & 0.6750  & 0.4673 & 0.1350  & 0.3637  & 0.2646 & 0.5340  & 0.3347 \\
AgentCF     & 0.2540 & 0.4210 & 0.3503 & 0.5470 & 0.4019 & 0.2160 & 0.5030 & 0.3802 & 0.6590 & 0.4443 \\
AgentCF++   & 0.2950  & 0.5243  & 0.4273 & 0.6737  & 0.4887 & 0.2073  & 0.4590  & 0.3504 & 0.6520  & 0.4292 \\
MemRec      & \underline{0.3610} & \underline{0.6440} & \underline{0.5280} & \underline{0.8010} & \underline{0.5926} & \underline{0.3520} & \underline{0.6240} & \underline{0.5113} & \underline{0.7620} & \underline{0.5685} \\
\method
            & \textbf{0.5410} & \textbf{0.8010} & \textbf{0.6968} & \textbf{0.8900} & \textbf{0.7344}
            & \textbf{0.3760} & \textbf{0.7740} & \textbf{0.5941} & \textbf{0.9040} & \textbf{0.6491} \\
\bottomrule
\end{tabular}
}
\end{table*}

\vspace{-6pt}
\subsection{Ablation Studies on Agent Actions (RQ2)}
\label{subsec:ablation}

To verify the effectiveness of the agent action design, we remove each core component from \method and report the results in Table~\ref{tab:rq2_ablation}. The ablation results show that the full model achieves the best or most reliable performance on almost all datasets and metrics. This indicates that the three sequential actions contribute complementary task-oriented evidence and jointly support the final recommendation decision.

Removing item semantic enhancement leads to a clear degradation, especially on Books and MovieTV. This suggests that the enhanced item semantics $d_i^E$ are important for constructing a task-aligned semantic anchor before the target user agent $A_u$ retrieves private-memory evidence $\mathcal{R}_t^U$. Without this anchor, the model has to rely more directly on the raw candidate description $d_i$, which may still preserve some high-recall behavior on certain datasets but provides a less precise basis for placing the ground-truth item at the top of the ranking list.

Without private-memory retrieval and projection, the target user agent retains only the user persona $\phi_u$ as user-side information for decision making. Consequently, the task-specific preference state $P_t$ loses the personalized evidence $\mathcal{R}_t^U$ needed to connect the current task with the user's historical preference semantics. The performance drop is consistent on Books, Goodreads, and MovieTV, indicating that evidence retrieved from $\mathcal{M}_u$ provides a stable source of personalized information across domains with informative interaction histories. On Yelp, although the variant achieves a higher H@1, the full model performs better on broader ranking metrics, indicating more reliable overall ranking quality.

Confidence-gated collaboration provides complementary evidence that cannot be fully recovered from the target user agent's private memory. Setting $\mathcal{R}_t^C=\varnothing$ consistently reduces performance across all four datasets, showing that collaborative user agents $A_v \in \mathcal N_u$ can supplement insufficient or ambiguous local evidence. The degradation is especially pronounced on Books and Yelp for different reasons. Specifically, Books combines sparse interactions with a large long-tail catalog, limiting the local memory's coverage of fine-grained reading tastes, whereas Yelp spans heterogeneous services and context-dependent attributes, making historical preferences less consistently aligned with the current request. 

\begin{table}
\centering
\caption{Ablation study on the four InstructRec datasets.}
\label{tab:rq2_ablation}
\setlength{\tabcolsep}{2.6pt}
\renewcommand{\arraystretch}{1.05}
\resizebox{\columnwidth}{!}{
\begin{tabular}{llccccc}
\toprule
\textbf{Dataset} & \textbf{Variant}
& \textbf{H@1} & \textbf{H@3} & \textbf{N@3} & \textbf{H@5} & \textbf{N@5} \\
\midrule

\multirow{4}{*}{Books}
& \textbf{\method}
& \textbf{0.5250} & \textbf{0.8173} & \textbf{0.6996} & \textbf{0.9327} & \textbf{0.7483} \\
\cmidrule(lr){2-7}
& w/o Semantic Enhancement
& 0.2523 & 0.5880 & 0.4477 & 0.7833 & 0.5288 \\
& w/o Confidence-Gated Collaboration
& 0.3933 & \underline{0.7283} & 0.5935 & \underline{0.8477} & 0.6439 \\
& w/o Private-Memory Retrieval
& \underline{0.4547} & 0.7003 & \underline{0.6022} & 0.8133 & \underline{0.6492} \\
\midrule

\multirow{4}{*}{Goodreads}
& \textbf{\method}
& \textbf{0.6860} & \textbf{0.9460} & \textbf{0.8480} & \textbf{0.9910} & \textbf{0.8668} \\
\cmidrule(lr){2-7}
& w/o Semantic Enhancement
& 0.5117 & \underline{0.9413} & \underline{0.7749} & \underline{0.9897} & \underline{0.7954} \\
& w/o Confidence-Gated Collaboration
& \underline{0.5853} & 0.8833 & 0.7668 & 0.9427 & 0.7915 \\
& w/o Private-Memory Retrieval
& 0.4893 & 0.7220 & 0.6301 & 0.8237 & 0.6724 \\
\midrule

\multirow{4}{*}{MovieTV}
& \textbf{\method}
& \textbf{0.5410} & \textbf{0.8010} & \textbf{0.6968} & \textbf{0.8900} & \textbf{0.7344} \\
\cmidrule(lr){2-7}
& w/o Semantic Enhancement
& 0.4213 & \underline{0.7923} & 0.6436 & \underline{0.8837} & 0.6818 \\
& w/o Confidence-Gated Collaboration
& \underline{0.5067} & 0.7820 & \underline{0.6706} & 0.8740 & \underline{0.7089} \\
& w/o Private-Memory Retrieval
& 0.4640 & 0.7030 & 0.6072 & 0.8107 & 0.6521 \\
\midrule

\multirow{4}{*}{Yelp}
& \textbf{\method}
& \underline{0.3760} & \textbf{0.7740} & \textbf{0.5941} & \underline{0.9040} & \textbf{0.6491} \\
\cmidrule(lr){2-7}
& w/o Semantic Enhancement
& 0.2577 & \underline{0.7503} & 0.5487 & \textbf{0.9100} & 0.6160 \\
& w/o Confidence-Gated Collaboration
& 0.2280 & 0.6950 & 0.5045 & 0.8653 & 0.5762 \\
& w/o Private-Memory Retrieval
& \textbf{0.4233} & 0.6923 & \underline{0.5840} & 0.8117 & \underline{0.6332} \\
\bottomrule
\end{tabular}
}
\end{table}

\vspace{-6pt}
\subsection{Sensitivity Analysis (RQ3)}

\begin{figure*}[t]
    \centering
    \resizebox{2\columnwidth}{!}{
        \includegraphics{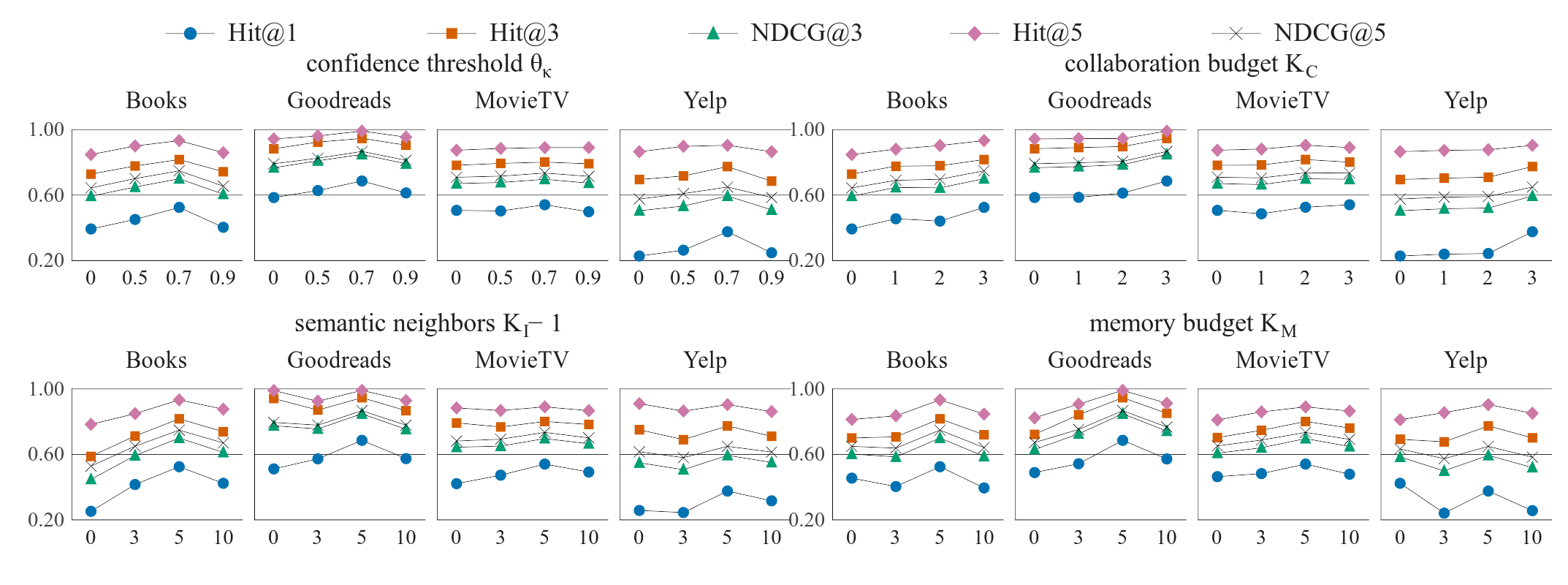}
    }
    \caption{Parameter sensitivity of \method on the four datasets with respect to the confidence threshold $\theta_\kappa$, collaboration budget $K_C$, number of semantic neighbors $K_I-1$, and target-user memory budget $K_M$. }
    \label{fig:paradigm_shift}
\end{figure*}

\subsubsection{Confidence Threshold $\theta_\kappa$ and Collaboration Budget $K_C$}
As shown in Fig.~\ref{fig:paradigm_shift}, both parameters indicate that collaboration is effective only within a controlled scope. A moderate $\theta_\kappa$ introduces external evidence for uncertain tasks while preserving reliable local decisions; a restrictive gate may miss useful cross-user signals, whereas a permissive gate admits weakly aligned patterns. The collaboration budget $K_C$ similarly exhibits saturation: expanding the collaborator set improves preference-pattern coverage until the main relevant signals are captured, after which additional agents become largely redundant or noisy. Thus, fragmented local histories may benefit from broader collaboration, while informative histories require fewer collaborators.

\vspace{-6pt}
\subsubsection{Platform-Local Item-Support Budget $K_I$}
The third panel of Fig.~\ref{fig:paradigm_shift} varies the semantic neighbors $K_I-1$, with the candidate included in the total budget $K_I$. Performance follows an inverted-U pattern: undersized support sets yield unstable enhancement, moderate neighborhoods produce the most reliable rankings, and excessive expansion degrades performance. With too few neighbors, incomplete or idiosyncratic candidate descriptions dominate, and the support is insufficient to reveal recurring task-relevant attributes. A moderate set establishes enough semantic agreement to calibrate the candidate, whereas broader neighborhoods introduce less aligned semantics that dilute its distinctive characteristics and cause semantic drift.
\vspace{-6pt}
\subsubsection{Target-User Memory Budget $K_M$}
Across datasets, $K_M$ shows a similar preference for concentrated evidence. Too few records leave $\mathcal{R}_t^U$ insufficient to ground $P_t$ beyond the user persona, while a moderate budget supplies enough aligned interactions to stabilize the task-specific preference state. Excessive retrieval instead introduces weakly related or contextually mismatched behaviors, diluting relevant signals and increasing the reasoning load. Occasional gains on individual metrics under small budgets do not persist across broader ranking measures, indicating that robust personalization depends on evidence relevance and concentration rather than memory volume.

\subsection{Effect of Agent Web Topology (RQ4)}

\begin{table}[t]
\centering
\caption{Effect of different user-agent edge-construction strategies in the Agent Web on the four InstructRec datasets.}
\label{tab:rq4_webgraph_construction}
\setlength{\tabcolsep}{2.6pt}
\renewcommand{\arraystretch}{1.05}
\resizebox{0.85\columnwidth}{!}{
\begin{tabular}{llccccc}
\toprule
\textbf{Dataset} & \textbf{Edge construction}
& \textbf{H@1} & \textbf{H@3} & \textbf{N@3} & \textbf{H@5} & \textbf{N@5} \\
\midrule

\multirow{3}{*}{Books}
& Item co-occurrence
& \textbf{0.5250} & \textbf{0.8173} & \textbf{0.6996} & \textbf{0.9327} & \textbf{0.7483} \\
& Persona-similarity
& \underline{0.4340} & \underline{0.7637} & \underline{0.6312} & \underline{0.8980} & \underline{0.6879} \\
& Random
& 0.3717 & 0.6620 & 0.5451 & 0.7857 & 0.5965 \\
\midrule

\multirow{3}{*}{Goodreads}
& Item co-occurrence
& \textbf{0.6860} & \underline{0.9460} & \textbf{0.8480} & \textbf{0.9910} & \textbf{0.8668} \\
& Persona-similarity
& \underline{0.6423} & \textbf{0.9517} & \underline{0.8328} & \underline{0.9840} & \underline{0.8466} \\
& Random
& 0.4910 & 0.7657 & 0.6559 & 0.8587 & 0.6945 \\
\midrule

\multirow{3}{*}{MovieTV}
& Item co-occurrence
& \textbf{0.5410} & \underline{0.8010} & \underline{0.6968} & \underline{0.8900} & \underline{0.7344} \\
& Persona-similarity
& \underline{0.5247} & \textbf{0.8230} & \textbf{0.7049} & \textbf{0.8977} & \textbf{0.7362} \\
& Random
& 0.4347 & 0.7113 & 0.5989 & 0.8130 & 0.6410 \\
\midrule

\multirow{3}{*}{Yelp}
& Item co-occurrence
& \textbf{0.3760} & \textbf{0.7740} & \textbf{0.5941} & \textbf{0.9040} & \textbf{0.6491} \\
& Persona-similarity
& \underline{0.2447} & \underline{0.7357} & \underline{0.5369} & \underline{0.9020} & \underline{0.6072} \\
& Random
& 0.2160 & 0.6030 & 0.4437 & 0.7993 & 0.5256 \\
\bottomrule
\end{tabular}
}
\vspace{-6pt}
\end{table}

To investigate how the Agent Web topology $\mathcal{G}=(\mathcal{A},\mathcal{E})$ affects task-time collaboration, we compare different strategies for constructing the user-agent edges in $\mathcal{E}$. The persona-similarity construction builds edges according to semantic similarity between persona embeddings, while the random construction preserves the out-degree of the item co-occurrence construction but randomly rewires its edges. This comparison examines whether task-time collaboration benefits from a scope induced by agent-level similarity.

Table~\ref{tab:rq4_webgraph_construction} shows that structured user-agent edge constructions consistently outperform random rewiring. Item co-occurrence achieves the strongest overall performance, suggesting that shared interaction histories provide a reliable signal for identifying collaborative user agents with homophilic interests. This verifies the role of the user-agent edges in $\mathcal{E}$ in determining the task-time collaboration scope: since not all collaborative evidence is useful for the target decision, preserving homophilic connectivity in the Agent Web helps acquire more relevant preference-pattern responses.

The persona-similarity construction also remains competitive, indicating that semantic user descriptions capture another useful form of agent-level proximity. However, its gains are less uniform than those of item co-occurrence, suggesting that high-level persona similarity may miss fine-grained preference evidence tied to the candidate item space. By contrast, the consistent degradation under degree-preserving random rewiring demonstrates that irrelevant agent links inject distracting evidence into collaboration, confirming that similarity-based connectivity is essential for preserving useful preference signals. Overall, these results show that different task-relevant similarity signals can be used to instantiate the user-agent edges in $\mathcal{E}$, but effective collaboration requires the Agent Web topology to encode task-relevant user proximity.

\vspace{-6pt}

%% file: relatedwork.tex
\section{Related Work}
\label{sec:related_work}

\subsection{Platform-Centric Recommendation}

Traditional recommenders follow a platform-centric pipeline that learns reusable user and item representations from centrally collected interactions. Matrix factorization, pairwise ranking, and neural collaborative filtering model global preference patterns~\cite{He2017NCF,Rendle2009BPR}, while sequential methods capture temporal dynamics and contextual dependencies~\cite{Kang2018SASRec,Sun2019BERT4Rec}. Their effectiveness relies on aggregating behavioral histories and optimizing a platform-wide ranking objective. Even when recent interests or high-order relations are modeled, the underlying evidence is still aggregated and interpreted centrally. Evidence processing and preference estimation therefore remain controlled by the platform, which differs from the Agent Web settings.

\vspace{-8pt}

\subsection{LLM and Agent-Based Recommendation}

LLM-based recommendation introduces language representations and semantic reasoning into recommendation. Existing methods formulate recommendation as instruction following, textual sequence modeling, or semantic item generation~\cite{Geng2022P5,Zhang2023InstructRec,Li2023Recformer,Rajput2023GenerativeRetrieval}, and incorporate collaborative semantics or augment sparse interaction graphs~\cite{LCRec2023,LLMRec2023}. Surveys further summarize the potential of LLMs for generative and general-purpose recommendation~\cite{Wu2023LLMRecSurvey,Li2023GenerativeRecSurvey}. Retrieval-augmented and graph-grounded approaches improve grounding by injecting documents, collaborative signals, or graph contexts into LLM reasoning~\cite{Gao2023RAGSurvey,Zhu2025CRAG,Qiu2025GCRS}. In most cases, however, retrieval remains a platform-side operation rather than an evidence-acquisition decision made by a user-owned agent.

LLM-powered agents extend this paradigm with planning, tool use, memory, reflection, and environmental interaction~\cite{Yao2023ReAct,Park2023GenerativeAgents,Wang2024AgentSurvey}. Recommendation agents can invoke tools, plan zero-shot procedures, coordinate specialized modules, and support interactive recommendation~\cite{Huang2023InteRecAgent,Wang2023RecMind,Wang2024MACRec}; other approaches simulate users or refine recommendations through agentic feedback loops~\cite{Zhang2023Agent4Rec,Zhang2023AgentCF,Cai2024AFL}. Memory-oriented studies further address long histories, reusable personas, lifelong preference modeling, and collaborative memory~\cite{Shi2025PersonaX,Zhu2024LIBER,Chen2026MemRec,Peng2025AgentRecSurvey}. Despite improving adaptability, these methods often wrap platform workflows or summarize histories into persistent contexts for a downstream LLM. \method instead treats recommendation as task-oriented evidence reasoning, grounding each decision in relevant agent-owned evidence across the Agent Web.

\vspace{-6pt}

%% file: conclusion.tex
\section{Conclusion}
\label{sec:conclusion}

This paper investigates personalized recommendation in the emerging \aw, where LLM-based personal agents maintain personalized context beyond individual platforms and collectively provide distributed user-side knowledge. Realizing this opportunity requires reasoning over opaque memories through bounded query-response interactions and aligning heterogeneous evidence with each decision. We formulate the problem as task-time evidence acquisition and progressive fusion, and propose \method, a framework that constructs compact decision evidence while keeping agent memories local. \method enhances candidate semantics with platform knowledge, grounds them in the target agent's relevant private experience, and, when local semantic evidence is insufficient, supplements the decision with generalizable preference patterns distilled from related user agents. Experiments on four InstructRec datasets demonstrate strong and consistent performance against baselines. Overall, our work establishes task-oriented evidence acquisition and progressive fusion as a principled foundation for personalized recommendation in the \aw.

%% file: appendix.tex

\newcounter{prompttemplate}
\renewcommand{\theprompttemplate}{\arabic{prompttemplate}}
\providecommand*{\prompttemplateautorefname}{Prompt Template}

\newcommand{\ptslot}[1]{\texttt{\{\{#1\}\}}}

\newenvironment{prompttemplate}[1]{%
  \par\medskip
  \refstepcounter{prompttemplate}%
  \hrule height 0.8pt
  \nobreak\vspace{2pt}
  \noindent
  \textbf{Prompt~\theprompttemplate}\enspace #1%
  \par\nobreak\vspace{2pt}
  \hrule height 0.4pt
  \nobreak\vspace{3pt}
  \small
  \raggedright
  \setlength{\parindent}{0pt}%
}{%
  \par\nobreak\vspace{2pt}
  \hrule height 0.4pt
  \par\medskip
}


\clearpage
\appendix

\section{Appendix}
\subsection{LLM Prompt Templates}
\label{sec:llm_prompts}

We instantiate each template as a single user-role message using the model's chat template, without a separate system prompt. Identically named placeholders refer to the same runtime variables across templates.

\setcounter{prompttemplate}{0}


\begin{prompttemplate}
  {\textsc{AbstractLLM}: Item semantic enhancement}
\label{fig:AbstractLLM}

You are an item-analysis agent in a recommendation system.\par
\smallskip

\textbf{\#\# Candidate item}\par
Description: \ptslot{candidate\_description}\par
Attributes: \ptslot{candidate\_attributes}\par
\smallskip

\textbf{\#\# Related items
(by semantic + attribute similarity)}\par
\ptslot{related\_items\_block}\par
\smallskip

\textbf{\#\# Task}\par
Enhance the semantic description of the candidate item by focusing on the COMMON patterns shared across the candidate and its related items.\par

Specifically:
\begin{itemize}
  \item Identify recurring semantic themes and attribute patterns among related items.
  \item Extract the shared characteristics that define this group of items.
  \item Use these shared patterns to refine and strengthen the description of the candidate item.
  \item If a text contains a [TRUNCATED: ...] marker, use both remaining content and attributes to infer key semantics.
\end{itemize}

Do NOT introduce unrelated details or emphasize differences.\par
Focus on pattern abstraction and semantic consolidation.\par
Return only the enhanced description as plain text.

\end{prompttemplate}

In this template, \ptslot{candidate\_description} and \\
\ptslot{candidate\_attributes} denote the raw item description $d_i$. Together, these two fields provide the content of the candidate item record.
\ptslot{related\_items\_block} contains the descriptions the related item memory records ${m^P_j}$ for $j\in\mathcal J_t^P(i)$. \textsc{AbstractLLM} extracts their recurring semantic and attribute patterns and returns the enhanced item semantics $d^E_i$.


\begin{prompttemplate}
  {\textsc{ProjectLLM}: Target-user preference projection}
\label{fig:ProjectLLM}

You are a user-modeling agent.\par
\smallskip

\textbf{\#\# User persona}\par
\ptslot{target\_user\_persona}\par
\smallskip

\textbf{\#\# Recent interaction memories}\par
\ptslot{retrieved\_target\_user\_memory\_snippets}\par
\smallskip

\textbf{\#\# Task}\par
Summarise the user's recent preferences in 2--3 sentences.\par
Return plain text only (no JSON, no markdown).

\end{prompttemplate}

Here, \ptslot{target\_user\_persona} corresponds to the target-user persona $\phi_u$, and \ptslot{retrieved\_target\_user\_memory\_snippets} contains the review representation of the retrieved memories $\widehat{\mathcal M}_{u,t}$. These memories provide the target-user evidence $\mathcal R_t^U$ relevant to the current recommendation task.

\textsc{ProjectLLM} summarizes this evidence as the task-specific preference state $P_t$. Its output is subsequently inserted into \\ \ptslot{target\_user\_preference\_summary}.


\begin{prompttemplate}
  {\textsc{MatchLLM}: Listwise local matching}
\label{fig:MatchLLM}

You are a recommendation decision agent.\par
You must output ONLY valid JSON.\par
\smallskip

\textbf{\#\# Candidate item list}\par
\ptslot{structured\_candidate\_item\_list}\par
\smallskip

\textbf{\#\# Candidate Item IDs (authoritative)}\par
\ptslot{candidate\_item\_id\_list}\par
Total candidates: \ptslot{num\_candidates}\par
\smallskip

\textbf{\#\# User preference summary}\par
\ptslot{target\_user\_preference\_summary}\par
\smallskip

\textbf{\#\# Task}\par
Score every candidate item against the user preference.
\begin{itemize}
  \item Use each item's `Item ID' as the unique key.
  \item score: float in [0, 1]
  \item confidence: float in [0, 1]
  \item rationale: one concise sentence
  \item If uncertain, still provide a conservative score; never skip.
\end{itemize}

\textbf{\#\# Hard constraints (must follow)}
\begin{itemize}
  \item Output EXACTLY one decision for each candidate Item ID above.
  \item Do NOT omit any candidate Item ID.
  \item Do NOT add extra Item IDs.
  \item Keep \texttt{item\_id} as integer.
  \item Ensure every decision has non-empty rationale.
  \item Before finalizing, verify decisions count equals total candidates.
\end{itemize}

Return JSON only in this format:\par
\texttt{\{"decisions": [}\par
\texttt{\{"item\_id": <int>, "score": <float>,}\par
\texttt{"confidence": <float>, "rationale": "<text>"\}]\}}

\end{prompttemplate}

In this template, \ptslot{structured\_candidate\_item\_list} contains the identifier, enhanced semantics $d_i^E$ of each candidate item. \ptslot{candidate\_item\_id\_list} provides the ordered candidate identifiers, and \ptslot{num\_candidates} provides their total number.  \\
\ptslot{target\_user\_preference\_summary} is the task-specific preference state $P_t$ produced by \textsc{ProjectLLM}. 
For each candidate, \textsc{MatchLLM} produces the local decision $D_t=(\hat{Y_t},\kappa_t,\rho_t)$, consisting of a matching score, confidence, and rationale. The complete JSON output is reused as \ptslot{list\_decisions\_json} in the subsequent templates.


\begin{prompttemplate}
  {\textsc{QueryLLM}: Collaboration subquery generation}
\label{fig:QueryLLM}

You are a central recommendation agent generating a subquery for
multi-agent collaboration in ranking.\par
\smallskip

\textbf{\#\# Current list decisions}\par
\ptslot{list\_decisions\_json}\par
\smallskip

\textbf{\#\# Task}\par
Collaboration has already been triggered. Based on the current list
decisions, generate one concise subquery that specifies the complementary
information required from neighbor agents.

\begin{itemize}
  \item The subquery should be a declarative sentence describing which
        item feature is uncertain and needs neighbor evidence.
  \item Focus on the uncertainty reflected in the current scores,
        confidence values, and rationales.
\end{itemize}

Return JSON only:\par
\texttt{\{"subquery": "}
\ptslot{collaboration\_subquery}
\texttt{"\}}

\end{prompttemplate}

Here, \ptslot{list\_decisions\_json} is the JSON output produced by
\textsc{MatchLLM}. It contains the preference score $\hat{Y_t}$, confidence $\kappa_t$,
and rationale $\rho_t$ for each candidate item.
\textsc{QueryLLM} uses the uncertainty reflected in these decisions to
construct the collaboration subquery $q_t^C$, represented by
\ptslot{collaboration\_subquery}. This prompt generates the information request after collaboration has been triggered.


\begin{prompttemplate}
  {\textsc{PatternLLM}: Collaborative pattern abstraction}
\label{fig:PatternLLM}

You are a neighbor user-agent in a recommendation graph.\par
\smallskip

\textbf{\#\# Subquery}\par
\ptslot{collaboration\_subquery}\par
\smallskip

\textbf{\#\# Your persona}\par
\ptslot{collaborating\_user\_persona}\par
\smallskip

\textbf{\#\# Matched memories (description + review)}\par
\ptslot{collaborator\_memory\_snippets}\par
\smallskip

\textbf{\#\# Task}\par
Use matched memories to answer the subquery with concise,
evidence-grounded text.\par

Return plain text only (no JSON, no markdown). Convey confidence through
your wording.

\end{prompttemplate}

In this template, \ptslot{collaboration\_subquery} is the same collaboration query $q_t^C$ returned by \textsc{QueryLLM}. \\
\ptslot{collaborating\_user\_persona} corresponds to the collaborator
persona $\phi_v$, while 
\ptslot{collaborator\_memory\_snippets} contains the
description review representation of the collaborator memories
$\widehat{\mathcal M}_{v,t}$. For each collaborator $A_v$,
\textsc{PatternLLM} uses these memories to produce an evidence-grounded
preference pattern $F_{v,t}^C$. The returned patterns are collected and
passed to \textsc{RefineLLM} through
\ptslot{neighbor\_responses\_with\_personas\_json}.


\begin{prompttemplate}
  {\textsc{RefineLLM}: Collaboration-based decision refinement}
\label{fig:RefineLLM}

You are the central agent that revises item-list ranking after one-hop
collaboration.\par
You must output ONLY valid JSON.\par
\smallskip

\textbf{\#\# Candidate item list}\par
\ptslot{structured\_candidate\_item\_list}\par
\smallskip

\textbf{\#\# Original list decisions}\par
\ptslot{list\_decisions\_json}\par
\smallskip

\textbf{\#\# Required Item IDs (authoritative)}\par
\ptslot{candidate\_item\_id\_list}\par
Total required decisions: \ptslot{num\_candidates}\par
\smallskip

\textbf{\#\# Neighbor responses (with persona)}\par
\ptslot{neighbor\_responses\_with\_personas\_json}\par
\smallskip

\textbf{\#\# Task}\par
Notice the recurring and mutually consistent signals from the neighbor responses and use neighbor responses and personas to revise the list decisions.
\begin{itemize}
  \item Keep \texttt{item\_id} unchanged.
  \item score/confidence in [0, 1].
  \item rationale should mention key evidence briefly.
  \item Treat item IDs in `Original list decisions' as authoritative.
  \item Output EXACTLY one decision for each original \texttt{item\_id}.
  \item Output the same \texttt{item\_id} set as Required Item IDs in the
        same order.
  \item Do NOT omit any original \texttt{item\_id}.
  \item Do NOT add any new \texttt{item\_id}.
  \item Do NOT output only top items; revise all required items.
  \item If neighbor evidence is irrelevant, keep the original decision.
  \item Keep each rationale to one concise sentence.
  \item Before finalizing, verify decision count equals original count.
\end{itemize}

Return JSON only:\par
\texttt{\{"decisions": [}\par
\texttt{\{"item\_id": <int>, "score": <float>,}\par
\texttt{"confidence": <float>, "rationale": "<text>"\}]\}}

\end{prompttemplate}

\textsc{RefineLLM} takes the candidate list and local decisions produced by \textsc{MatchLLM}, together with the collaborator evidence returned by \textsc{PatternLLM}. It uses this additional evidence which recurs across collaborators to update the local decision for each candidate while preserving the original candidate set and order. 